\documentclass[floats,epsf,showpacs]{elsarticle}
\usepackage{graphicx}
\usepackage{epstopdf}
\usepackage{hyperref}
\usepackage{amsmath}

\usepackage{color}

\newcommand{\nop}[1]{}
\def\pM{\ensuremath{\genfrac{}{}{0pt}{1}{+}{\scriptstyle(\kern-1pt-\kern-1pt)}}}

\journal{Journal of Magnetism and Magnetic Materials}

\begin{document}

\begin{frontmatter}

\title{Dynamical vertex approximation for the two-dimensional Hubbard model}

%% Group authors per affiliation:
\author{Thomas Sch\"afer}
\author{Alessandro Toschi}
\author{Karsten Held}
\ead{held@ifp.tuwien.ac.at}
\address{Institute for Solid State Physics, TU Wien, 1040 Vienna, Austria}

\begin{abstract}
Recently,  diagrammatic extensions of dynamical mean field theory (DMFT) have been proposed for including short- and long-range correlations beyond DMFT on an equal footing. We employ one of these, the dynamical vertex approximation (D$\Gamma$A), and study the two-dimensional Hubbard model on a square lattice. We define two transition lines in the phase diagram which correspond, respectively, to the opening of the gap in the nodal direction and throughout the Fermi surface. 
Our self-energy data show that the evolution between the two regimes occurs in a gradual way (crossover) and also that at low enough temperatures the whole Fermi surface is always gapped. 
Furthermore, we present a comparison of our D$\Gamma$A calculations at a parameter set where data obtained by other techniques are available.
\end{abstract}

\begin{keyword}
Strongly correlated electron systems, Mott-Hubbard transition
PACS: 71.27.+a, 71.30.+h
\end{keyword}

\end{frontmatter}

  \section{Introduction}
Dynamical mean field theory (DMFT)
 \cite{DMFT1,DMFT2,DMFT3,PhysToday} has been a big step forward for 
the calculation and understanding of strongly correlated electron systems. 
It includes a major part of the electronic correlations: the local ones. 
In the vicinity of phase transitions or in one- or two dimensions non-local correlations are however essential. To include non-local correlations but to keep the local DMFT correlations at the same time, cluster extensions of DMFT have been
developed such as the
dynamical cluster approximation (DCA) \cite{Maier04,LichtensteinDCA} and cluster DMFT \cite{clusterDMFT,LichtensteinDCA}. However numerical restrictions regarding the size of the clusters only allow to treat short range correlations this way.

Hence, as an alternative and to deal with short- and long-range correlations 
on the same footing, diagrammatic vertex extensions of DMFT have been proposed more recently. 
The first such approach has been the  dynamical vertex approximation 
(D$\rm \Gamma$A) \cite{DGA1}, followed by the  dual fermion (DF) approach  \cite{DualFermion}, the one-particle irreducible approach (1PI) \cite{1PI},
and the merger of DMFT with the functional renormalization group (DMF$^2$RG)  \cite{DMF2RG}.
All of these approaches start with a local two-particle vertex \cite{vertex,RohringerPhD} 
and calculate  from this non-local correlations beyond DMFT. This way, non-local correlations and associated 
physics are obtained, both, on the two-particle level as e.g. for the susceptibilities and on the one-particle-level as e.g. for the momentum-dependence of the self energy.
The difference between the various approaches lies  in which two-particle vertex is taken: the fully irreducible one (full D$\Gamma$A), the irreducible one in a certain channel (ladder   D$\Gamma$A), the one-particle irreducible  (1PI and DMF$^2$RG) or the reducible one (DF). Then, depending on the approach, Feynman diagrams are constructed from the
 full Green function  $G$ or  from the difference 
between  $G$ and the local Green function $G_{\rm loc}$; and different
kind of diagrams are taken: parquet, ladder, 2nd order or the ones generated by an RG flow.

Hitherto, the diagrammatic extensions of DMFT have been applied to simple models, in particular the one-band Hubbard model, though the concept of {\em ab initio} calculations with  D$\Gamma$A has also been proposed \cite{AbinitioDGA}.
Physical highlights so-far have been the calculation of the critical exponents of the three-dimensional Hubbard model \cite{Rohringer11,Hirschmeier2015} and the Falicov-Kimball model \cite{Antipov14}, quantum criticality \cite{Schaefer15}, pseudogaps physics \cite{Katanin,Rubtsov2009}, and superconductivity \cite{Otsuki14}. It was also possible to show that the paramagnetic phase of the square lattice Hubbard model is always insulating at low enough temperatures $T$ \cite{Schaefer14}, 
i.e. that the whole metallic side of the Mott-Hubbard metal-to-insulator transition as described by DMFT is completely washed out by extended long-range spin fluctuations in 2D.

In this paper we recapitulate the D$\Gamma$A method in Section 
\ref{Sec:method}; for further details the reader is referred to \cite{RohringerPhD} and \cite{HeldJuelich}. Results for the two-dimensional Hubbard model are presented in Section 
\ref{Sec:results}. Beyond  \cite{Schaefer14}, we analyze further aspects of the transition from the high-$T$ paramagnetic metallic phase to the low-$T$ paramagnetic insulator. In particular, we start by identifying two crossover lines in the phase diagram: 
first the self-energy  at the antinodal 
 ${\mathbf k}=(\pi,0)$ turns insulating; at a lower $T$ also nodal momentum ${\mathbf k}=(\pi/2,\pi/2)$ shows an insulating behavior.
Specifically, our self-energy data demonstrates that the evolution between the two regimes takes place {\em gradually} by decreasing $T$ for all the $U$ values analyzed.
In Section \ref{Sec:comparison}, we also present, as a benchmark, the  comparison of 
 D$\Gamma$A with DF and DCA. Section \ref{Sec:conclusion} summarizes our results.

\section{Dynamical vertex approximation}
\label{Sec:method}

\begin{figure}[tb]

 \centering \includegraphics[width=\textwidth]{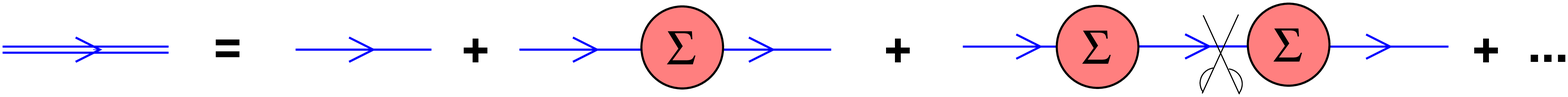}
 \caption{Graphical representation of the Dyson equation.\label{Fig:dyson}}
%, connecting the Green function $G$ (double blue line), its non-interacting counterpart  $G_0$ (single blue line) and the self energy $\Sigma$ (red circles).  From the irreducible self energy, reducible Green diagrams are generated, i.e., cutting a Green function line seperates the diagram into two pieces as indicated by the pair of scissors.\label{Fig:dyson}}
\end{figure}

The idea of D$\Gamma$A is a resummation of Feynman diagrams in terms of the locality of the diagrams, not in orders of $U$. 
The first, one-particle level of this resummation is DMFT,
which approximates the self energy, which is the one-particle 
fully irreducible vertex, to be local. Here, irreducible means that cutting one Green function line does not separate any self energy diagram into two pieces. Such reducible contributions are then generated by the Dyson Eq.:
\begin{equation}
G(\nu,{\mathbf k})=[1/G_0(\nu,{\mathbf k})-\Sigma(\nu,{\mathbf k})]^{-1} ,
\label{Eq:Dyson}
\end{equation}
 connecting the Green function $G(\nu,{\mathbf k})$ at Matsubara frequency $\nu$ 
and momentum $k$ (double blue line in Fig.\ \ref{Fig:dyson}), its non-interacting counterpart   $G_0(\nu,{\mathbf k})=1/[\nu+\mu-\epsilon_k-\Sigma(\nu,{\mathbf k})]$ (single blue line) and the self energy $\Sigma(\nu,{\mathbf k})$ (red circles).  %From the irreducible self energy
%That is, from the irreducible self energy $\Sigma$ at Matsubara frequeny $\nu$ 
%and momentum $k$ and the non-interacting Green function $G_0(\nu,{\mathbf k})=1/[\nu+\mu-\epsilon_k-\Sigma(\nu,{\mathbf k})]$, the Dyson equation generates all (reducible and irrediucible) diagrams for the interacting Green function 
% $G$, see Fig.\ \ref{Fig:dyson}.

The next, two-particle level then assumes the two-particle fully irreducible vertex to be local. This is D$\Gamma$A as employed nowadays. Before we discuss this approach in further detail, let us mention that, in principle, the fully irreducible $n$-particle vertex can be assumed to be local. This way more and more contributions are generated, and for $n\rightarrow \infty$ the exact solution is recovered. However with increasing $n$ the numerical effort also explodes, and going beyond $n=2$, or possibly $n=3$, does not seem to be feasible.

Let us hence now focus on the two-particle level. The full two-particle vertex $F$
consists of all connected diagrams. Some of these diagrams are two-particle fully irreducible, i.e., cutting two Green function lines $G$
does not separate the diagram into two pieces. We denote these fully irreducible diagrams by $\Lambda$. Other (reducible)  diagrams  separate into two pieces. 
(Here and in the following we consider skeleton diagrams, i.e., Feynman diagrams in
 terms of
the interacting Green function  $G$ instead of those in terms of $G_0$.)

\begin{figure}[tb]
 \includegraphics[width=\textwidth]{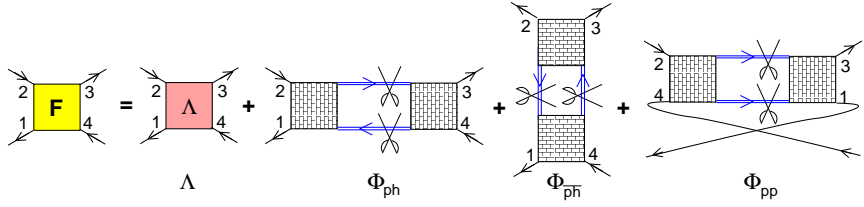}
 \caption{Graphical representation of the parquet equation:
the full  vertex $F$ contains fully two-particle irreducible diagrams
 $\Lambda$ and two-particle reducible diagrams in the 
particle-hole $\Phi_{ph}$, transversal particle-hole $\Phi_{\overline{ph}}$,
and  particle-particle reducible  channel $\Phi_{{pp}}$. \label{Fig:DGAF5}}
\end{figure}

The {\em reducible diagrams} of $F$ can be further classified.
If we cut two  $G$ lines, there are three ways how 
 $F$  separates into two pieces: particle-hole ($\Phi_{ph}$),
 particle-hole transversal ($\Phi_{\overline{ph}}$) and particle-particle ($\Phi_{pp}$)  reducible diagrams, see
 Fig.~\ref{Fig:DGAF5}. One can prove, using particle-conservation, that 
each diagram is actually either fully irreducible or reducible in {\em exactly}
one channel, so that
\begin{equation}
F(12 34)=\Lambda(1234)+\Phi_{ph}(1234)+\Phi_{\overline{ph}}(1234)+\Phi_{{pp}}(1234).\label{Eq:F1}
\end{equation}
Here, we have introduced a short-hand notation: 
$1$ represents a momentum-frequency-spin coordinate $1 \equiv ({\mathbf k},\nu,\sigma)$, $2 \equiv ({\mathbf k + \mathbf q},\nu+\omega,\sigma)$, $3\equiv ({\mathbf k' + \mathbf q},\nu'+\omega,\sigma')$,
$4 \equiv ({\mathbf k'},\nu',\sigma')$.
Eq.\ (\ref{Eq:F1}) is known as   the {\em parquet equation}.

Eq.\ (\ref{Eq:F1}) has, for a given  $\Lambda$, four unknowns: $F$ and
three $\Phi_r$'s $r \in\{ ph, \overline{ph}, pp\}$.
Next, we hence first define  the irreducible diagrams in a certain channel $r$:
\begin{equation}
         \Gamma_r(1234) \equiv F(1234)- \Phi_{r}(1234) \; .
\label{Eq:F2}
\end{equation}
From   $\Gamma_r$, we can then calculate the reducible diagrams 
by repeatedly connecting them by two Green functions, i.e., formally
 $\Phi_{r}  =\Gamma_r  GG  \Gamma_r +   \Gamma_r  GG \Gamma_r GG \Gamma_r + \dots$. Inserting this as a recursion into Eq. (\ref{Eq:F2}), yields the 
 {\em Bethe-Salpeter Eqs.} in the three channels (using Einstein's summation convention), cf.\ Fig.\ \ref{Eq:BS3}:
\begin{eqnarray}
F(1234)&=&\Gamma_{ph}(1234) + F(122'1')  G(3'2') G(1'4') \Gamma_{\rm ph}(4'3'34)\label{Eq:BS1}\\ 
&=&\Gamma_{\overline{ph}}(1234) +  F(2'233') G(2'1') G(3'4')\Gamma_{\overline{ph}}(11'4'4)\label{Eq:BS2}\\
&=&\Gamma_{{pp}}(1234) +  F(4'22'4)  G(2'3') G(1'4')\Gamma_{{pp}}(13'31').
\label{Eq:BS3}
\end{eqnarray}
If we now substitute    $\Gamma_r(1234) \equiv F(1234)- \Phi_{r}(1234)$ in the three
Bethe-Salpeter equations, we have four 
equations [Eqs.\ (\ref{Eq:F1}), (\ref{Eq:BS1}), (\ref{Eq:BS2}),  (\ref{Eq:BS3})],  which can be resolved for the four unknowns [$F$, $\Phi_{ph}$,  $\Phi_{\overline{ph}}$, and $\Phi_{pp}$].

\begin{figure}[tb]
 \centering \includegraphics[width=0.6\textwidth]{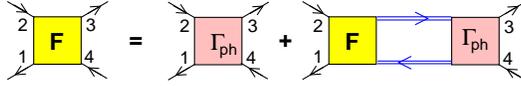}
 \caption{Bethe-Salpeter Eq. (\ref{Eq:BS1}) in the particle-hole ($ph$) channel.
 \label{Fig:DGAF6}}
\end{figure}

There is however 
one further complication, namely the internal Green function lines 
$G$ in the  Bethe-Salpeter Eqs. These  $G$'s can be calculated from $\Sigma$ via the Dyson Eq. (\ref{Eq:Dyson}); and $\Sigma$ in turn can be calculated exactly  from $F$ 
through the   Heisenberg
Eq. of motion (also called Schwinger-Dyson Eq. in this context):
\begin{eqnarray}
\Sigma(14) &=& - U(12'3'1')  G(1'4') G(23') G(2'3) F(4'234) \nonumber \\ && + U(1234)G(23) -  U(1432)G(23) 
\label{Eq:eqofmotion}
\end{eqnarray} 

%\begin{figure}[tb]
% \centering \includegraphics[width=.8\textwidth]{FigDGAF8.eps}
% \caption{Graphical represnetation of the Schwinger-Dyson equation which allows to calculate  $\Sigma$ from $U$, $F$ and $G$.
%Note, the last two diagrams  are the Hartree and Fock terms not included in $F$. \label{Fig:DGAF8}}
%\end{figure}

That is the solution of the exact parquet Eqs. involves
six equations  [(\ref{Eq:F1}, (\ref{Eq:BS1}), (\ref{Eq:BS2}),  (\ref{Eq:BS3}), (\ref{Eq:eqofmotion}), (\ref{Eq:Dyson})] and six unknowns [$F$, $\Phi_{ph}$,  $\Phi_{\overline{ph}}$, $\Phi_{pp}$, $\Sigma$, $G$]. If the exact $\Lambda$ were known, 
this would allow us to calculate the exact vertex $F$,  $\Sigma$, and $G$.
If not, we still can do approximations for $\Lambda$. The so-called parquet approximation assumes $\Lambda=U$ \cite{Bickers}; in D$\Gamma$A we consider all Feynman diagrams  for  $\Lambda$ in all orders of $U$, but only take their local contribution.
Let us note that the locality assumption for  $\Lambda$ is much better fulfilled
than that for $\Sigma$. Even for the two dimensional model considered in this paper. DCA calculations have shown  $\Lambda$ to be essentially ${\mathbf k}$-independent \cite{Maier06} while $\Sigma$ and $F$ are  strongly  ${\mathbf k}$-dependent.

So-far we have discussed the full (or parquet) D$\Gamma$A.
 Often  (also for the half-filled Hubbard model discussed in this paper) spin-fluctuations dominate over charge fluctuations  at low energies. 
In this case, the particle-hole and particle-hole transversal channels dominate over the  particle-particle channel. Hence, we can neglect $\Phi_{pp}$.
Then, the  $\Phi_{ph}$ and $\Phi_{\overline{ph}}$ channels decouple, and
are even related by crossing-symmetry \cite{RohringerPhD}, so that the calculation of the two Bethe-Salpeter Eqs. (\ref{Eq:BS1}), (\ref{Eq:BS3}) (or even one because of the crossing symmetry with a local  $\Gamma_{ph}$ as a starting point) is sufficient. This ladder-D$\Gamma$A is numerically much easier and
the two channels modulo the double counting can be combined  to yield the full vertex in Eq.\ (\ref{Eq:F1})   \cite{RohringerPhD,Katanin}.

\section{Metal-insulator crossover in the 2D Hubbard model} \label{Sec:results}

Let us   now turn to the ladder-D$\Gamma$A results for the half-filled Hubbard model on a square lattice with nearest neighbor hopping $t$ and local Coulomb repulsion $U$, given by the Hamiltonian
\begin{equation}
 H=-t\sum\limits_{\langle ij \rangle \sigma}{c^{\dagger}_{i\sigma}c^{\phantom\dagger}_{j\sigma}}+U\sum\limits_{i}{n_{i\uparrow}n_{i\downarrow}}.
 \label{eqn:hubb}
\end{equation}
Here, $c^{\dagger}_{i\sigma}$ ($c^{\phantom\dagger}_{i\sigma}$) creates (annihilates) an
electron on lattice site $i$ with spin $\sigma$, $\langle ij \rangle$ sums each nearest neighbor pair once, and energies are from now on measured in units of the half-bandwidth $D=4t$.

The main physical result of \cite{Schaefer14} is that the
low-temperature paramagnetic phase (above the $T=0$ antiferromagnet) is always insulating 
 for all $U>0$. In other words, the metal-insulator transition is at $U_c=0$ instead of a finite value $U_c>0$ which was previously concluded from  cluster extensions of DMFT, see, e.g., \cite{MIT_CDMFT}.
As a function of temperature $T$ however, there is a {\em crossover} from the low-$T$ paramagnetic insulator to a high-$T$ paramagnetic metal at small $U$'s.
 Here, we analyze in greater detail how this crossover occurs.

\begin{figure*}[tb]
    \centering
    \includegraphics[width=1\textwidth]{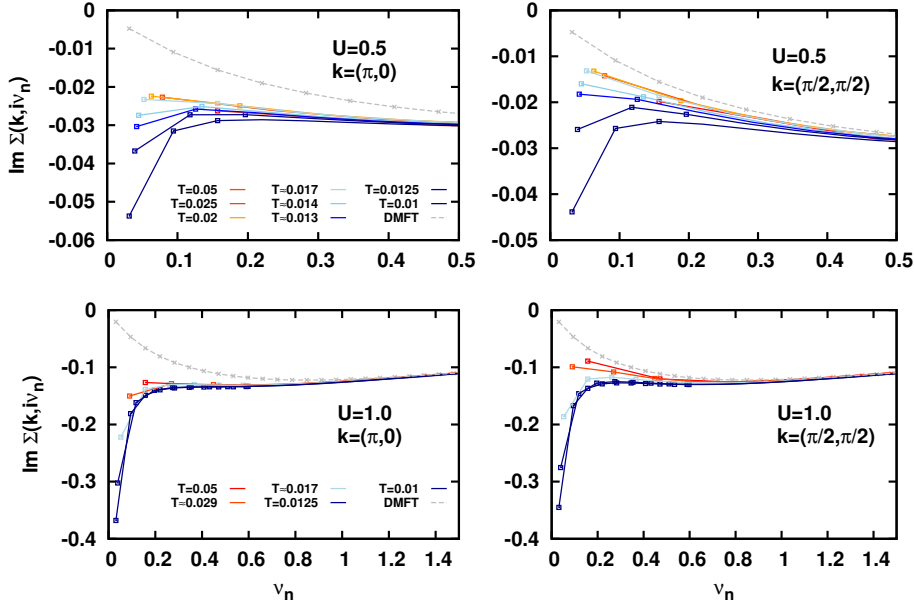}
    \caption{Imaginary parts of the D$\Gamma$A self-energy $\Sigma$ vs Matsubara frequency $\nu_n$ for the half-filled Hubbard model at the antinodal (left) and nodal (right) point of the Fermi surface at $U=0.5$ (upper panels) and $U=1.0$ (lower panels) and different temperatures. The DMFT results at   $T=0.01$ is provided for comparison.}
    \label{fig:self_fate}
\end{figure*}

 Fig. \ref{fig:self_fate} shows the  imaginary part of the self-energy for the nodal and antinodal points of the Fermi surface. For $U=0.5$ and high temperatures, $T>{0.017}$, 
the self-energy shows clearly metallic behavior at  every point on the Fermi surface
from the nodal to the antinodal one. Upon cooling along the  gray arrow  in Fig. \ref{fig:self_fate}, first the antinodal  point of the Fermi surface [$k=(\pi,0)$] shows a downturn for $\nu\rightarrow0$ in  Fig. \ref{fig:self_fate}, i.e., an insulating behavior below  $T={0.017}$. 

For  $0.017>T\geq 0.0125$ however, the nodal point of the Fermi surface [$k=(\pi/2,\pi/2)$] does not show this insulating behavior yet. Only for  $T<0.0125$ this and all other points of the Fermi surface show an insulating self energy.
In other words, for   $T<0.0125$ we have an insulator with the whole Fermi surface gapped,
whereas for  $0.017>T\geq 0.0125$ we have a pseudogap with only parts of the Fermi surface gapped.

\begin{figure*}[tb]
    \centering
    \includegraphics[width=0.6\textwidth]{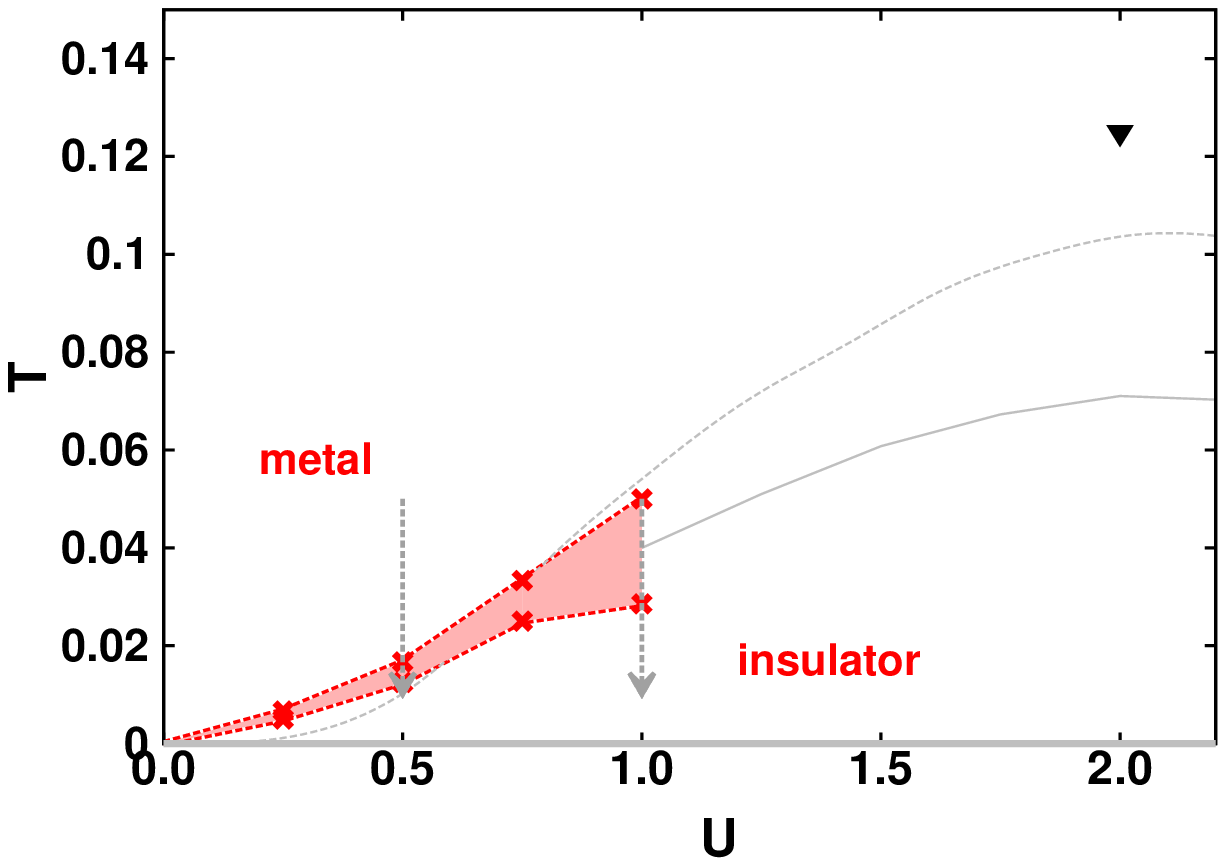}
    \caption{D$\Gamma$A phase diagram for the half-filled Hubbard model on a square lattice. 
Due to strong  antiferromagnetic fluctuations, we have a paramagnetic insulator for all $U>0$ below the solid-red line \cite{Schaefer14}: upon cooling (e.g. along the vertical arrows that mark the
temperatures presented in Fig.\ \ref{fig:self_fate})  first the antinodal point  [$k=(\pi,0)$]
turns insulating at the dashed-red line where the other $k$ points still show a metallic self energy. That is, in the red-shaded region, we have a crossover with a pseudogap. The dashed-gray and solid-gray line indicates the N\'eel temperature $T_N$ in DMFT and D$\Gamma$A for  3D, respectively (for 2D  $T_N=0$ in D$\Gamma$A). The triangle marks the parameters for the comparison in Fig.\ \ref{fig:siw_comp}.}
    \label{fig:pd}
\end{figure*}

Hence, upon cooling along the  grey arrow in the phase diagram Fig. \ref{fig:pd}, 
we hence first cross  the dashed red line which marks the temperature where the first point of the Fermi surface [$k=(\pi,0)$] shows insulating behavior. At lower temperatures, we  cross a second (solid-red) line at which the 
whole Fermi surface gets gapped. The red-shaded region inbetween hence has a  pseudogap.

The  analogous analysis for $U=1.0$ can be found in the lower panels of  Fig.\  \ref{fig:self_fate}. It  corresponds to the second gray arrow in  Fig. \ref{fig:pd}. In this case,
we identify a larger pseudogap region at higher temperatures: for $0.05>T>0.025$ 
the nodal self energy is still metallic whereas the antinodal one is already insulating.

\section{Comparison of numerical techniques for the 2D Hubbard model}\label{Sec:comparison}
\begin{figure*}[tb]
    \centering
    \includegraphics[width=0.75\textwidth]{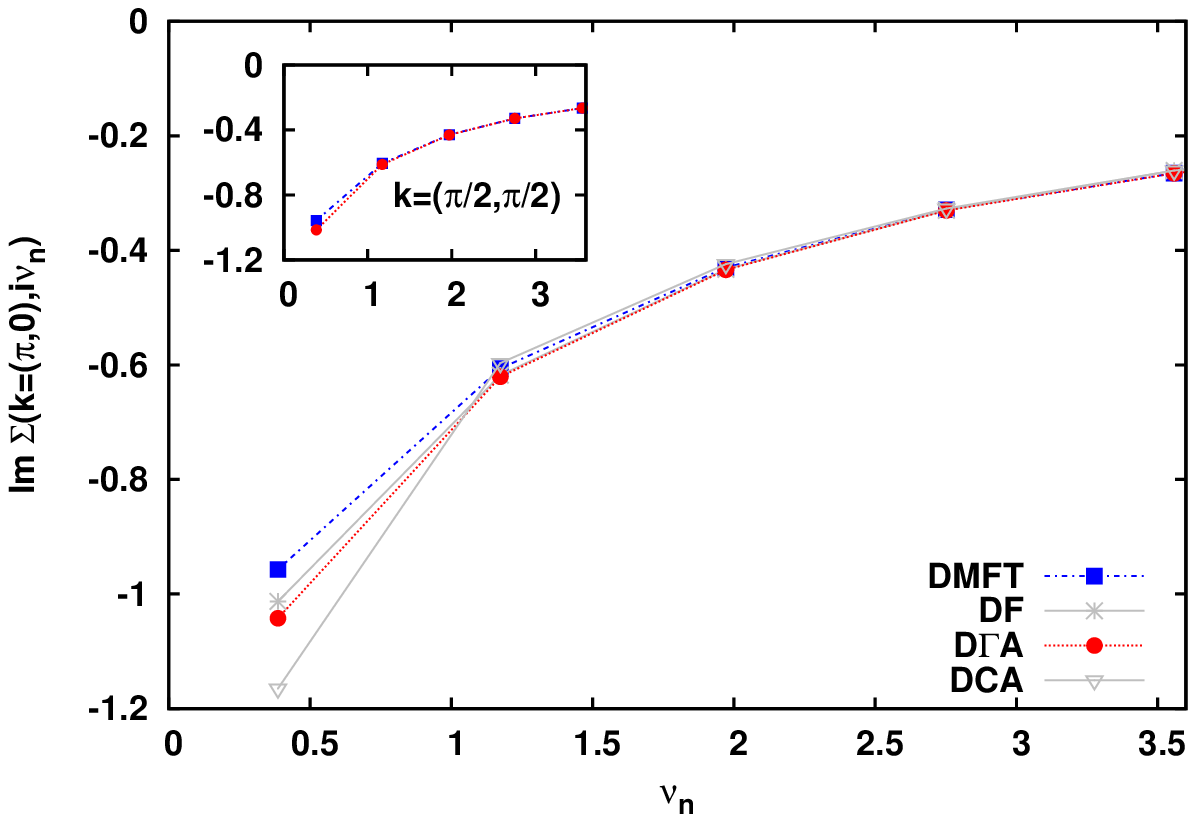}
    \caption{Comparison of the imaginary parts of DMFT (blue squares), D$\Gamma$A (red dots), DCA (grey triangles from \cite{LeBlanc15}) and DF (grey stars from \cite{LeBlanc15}) self-energies for the half-filled 2D Hubbard model with $U=2.0$ and $T=0.125$. The main plot shows the antinodal point ($k=(\pi,0)$) and the inset the nodal point ($k=(\pi/2,\pi/2)$) of 
the Fermi surface. For the latter, no DF/DCA data are available for comparison.}
    \label{fig:siw_comp}
\end{figure*}

As mentioned in the Introduction, for the treatment of the two-dimensional Hubbard model there exists a variety of numerical techniques, that aim at including correlations 
beyond the purely temporal but local ones covered by DMFT. Since there is no exact solution of this model in 2D, benchmarks between different methods are of fundamental importance. Fig. \ref{fig:siw_comp} shows 
a self-energy comparison of our DMFT and ladder-D$\Gamma$A  data with the DCA and DF results extracted from the extensive numerical review  \cite{LeBlanc15} by J. LeBlanc et al. for
 $U=2.0$ and $T=0.125$. These parameters correspond to the black triangle in the upper right corner of Fig. \ref{fig:pd}. 

In the main panel of  Fig. \ref{fig:siw_comp} one can see the data for the antinodal point $k=(\pi,0)$. Because of the high temperature and large interaction all data sets display an insulating behavior, also  DMFT.  However, the inclusion of non-local correlations (DCA, DF, D$\Gamma$A) further enhances the insulating tendencies; the self-energy downturn is more pronounced. The magnitude of this effect is similar in DF and D$\Gamma$A and somewhat larger in  DCA.  The overall trend is similar  for the nodal point  $k=(\pi/2,\pi/2)$ shown as the inset of  Fig. \ref{fig:siw_comp}. In contrast to the weak coupling region which we focused on  in
Section \ref{Sec:results}, for this larger $U$-value we have a Mott-Hubbard insulating behavior for all $k$ points of the Fermi surface. This reflects the increased importance of local correlations in this regime; even DMFT yields resonable results. 
% which does not capture spatial correlations (in this case extended antiferromagnetic ones), 
%&exhibits an insulating behavior for these parameters, which gets further pronounced including short-ranged (DCA) and long-ranged fluctuations (DF, D$\Gamma$A). However, as the parameter regime (high temperature, strong coupling, not 
%in the immediate vicinity of a phase transition) suggests, all the methods obtain similar results that only deviate in the the renormalization of the first and second Matsubara frequency.

\section{Conclusion} \label{Sec:conclusion}
Due to long-ranged antiferromagnetic correlations, the paramagnetic phase of
the two-dimensional Hubbard model is  insulating at low enough temperatures, half-filling, and
 a lattice with perfect nesting
such as the square lattice with nearest-neighbor hopping --- even at an arbitrary small coupling $U$ \cite{Schaefer14}. For small $U$ and higher temperatures, on the other hand, we have a paramagnetic
metal. In this paper, we analyzed the metal-to-insulator crossover in detail.
We find that upon cooling, first a gap opens at the antinodal point 
before, at a lower temperature, the full Fermi surface is gapped. That is, there is  a pseudogap.
We have also compared  our D$\Gamma$A results with DF and QMC where these are available, i.e.,  at  higher temperatures and intermediate coupling strength. In this parameter regime, DMFT already gives a good description, and all of the three beyond-DMFT approaches show similar corrections to DMFT: non-local correlations systematically increase the insulating tendencies.

{\em Acknowledgments.} We thank  M.~Aichhorn, E.~Arrigoni, N.~Bl\"umer, F.~Geles, E.~Gull, A.~Katanin, J.~P.~F.~LeBlanc, G.~Rohringer, D.~Rost, C.~Taranto, P.~Thunstr\"om, 
J.~M.~Tomczak, and A.~Valli for discussions. Financial support is acknowledged from the Austrian Science Fund (FWF) through graduate school W1243 Solid4Fun (T.S.) 
and SFB ViCoM (F41, A.T.), as well as from the European Research Council under the European Union's Seventh Framework Programme (FP/2007-2013)/ERC through grant agreement n. 306447 ({\em AbinitioD$\rm \Gamma$A}, K.H.). The calculations were performed on the Vienna Scientific Cluster (VSC).

%%%%%%%%%%%%%%% References %%%%%%%%%%%%%%%%%

\end{document}